\documentstyle[amssymb,12pt]{article}

\input tcilatex
\if@twoside
\else
\fi
\advance\textheight by \topskip
\leftmargin\leftmargini
\labelwidth\leftmargini\advance\labelwidth-\labelsep

\begin{document}

\begin{center}
\bigskip {\LARGE Duffin-Kemmer-Petiau Theory in the Causal Approach}

\vskip0.5cm

{\large J. T. Lunardi}$^{a}$\footnote{%
On leave from Universidade Estadual de Ponta Grossa, Setor de Ci\^{e}ncias
Exatas e Naturais, Departamento de Matem\'{a}tica e Estat\'{\i}stica. Ponta
Grossa-PR, Brazil.}{\large , L. A. Manzoni}$^{b}${\large , B. M. Pimentel}$%
^{a}$ {\large and}

{\large \ J. S. Valverde}$^{a}$

\vskip0.5cm

$^{a}$Instituto de F\'{\i}sica Te\'{o}rica

Universidade Estadual Paulista\\[0pt]
Rua Pamplona 145

01405-900 - S\~{a}o Paulo, SP

Brazil\\[0pt]

$^{b}$Instituto de F\'{\i}sica

Universidade de S\~{a}o Paulo

Caixa Postal 66318

05315-970 - S\~{a}o Paulo - SP

Brazil.
\end{center}

\vskip1.0cm

\begin{center}
\begin{minipage}{14.5cm}
\centerline{\bf Abstract}
{In this paper we consider the scalar sector of Duffin-Kemmer-Petiau 
theory in the framework of Epstein-Glaser causal method. We
calculate the lowest order distributions for Compton scattering, vacuum 
polarization, self-energy and vertex corrections. By requiring gauge invariance 
of the theory we recover, in a natural way, the scalar propagator of the 
usual effective theory.}
\end{minipage}
\end{center}

\vskip1.0cm

\section{Introduction}

The usual way to approach scalar quantum electrodynamics (SQED) is by
performing the electromagnetic minimal coupling in the free Lagrangian of
Klein-Gordon (KG) scalar field theory \cite{Itzikson}. An alternative way is
to start from the free Duffin-Kemmer-Petiau (DKP) Lagrangian instead of the
KG one.

The free DKP theory is a theory for scalar and vector fields \cite{Petiau,
Kemmer 1, Duffin, Kemmer 2, Umezawa} given by a Lagrangian formally similar
to that of spinorial QED. The fundamental differences are the algebraic
relations satisfied by the $\beta ^{\mu }$ matrices in DKP theory, which
play the role analogous to $\gamma ^{\mu }$ in spinor QED.

It is known that in the free field case the DKP and KG theories are
equivalent, both in classical and quantum pictures \cite{Itzikson,
Bogoliubov, Pimentel 2, Berestetskii}. However, there are still no general
proofs of equivalence between these theories when interactions and decays of
unstable particles are taken into account \cite{Krajcik, Pimentel 3} (in
this context see also references \cite{prl, prc1}, which relate different
results by using both DKP and KG formalisms with strong interactions). Some
progress in this direction has been made recently. For instance, it was
shown that both theories are equivalent in the classical level for the cases
of minimal interaction with electromagnetic \cite{Pimentel 2, Nowakowski}
and gravitational \cite{DKP curvo} fields . Strict proofs of equivalence
between both theories were also given for the cases of interaction of the
quantized scalar field with classical and quantized electromagnetic,
Yang-Mills and external gravitational fields \cite{Pimentel 3, Pimentel 1}.

Perhaps one of the most evident advantages in working with this theory is
the fact that derivative couplings do not appear between DKP and the gauge
field (this property has been used by Gribov recently, who employed the
vector sector of \ DKP theory to study the quark confinement problem \cite
{Gribov}). Such property will result in manifestly covariant expressions for
the interaction Hamiltonian and the vacuum expectation values of time
ordered products of fields. Another advantages are the formal similarity
with spinor QED (which facilitates the adaptation to the scalar case of
technics developed formerly in the spinor case \cite{prl, prog}) and the
fact that this formalism allows \ an unified treatment of the scalar and
vector fields.

One of the difficulties in working with SQED based on KG equation (SQED-KG)
is the presence of a term of second order in the coupling constant in the
interaction Hamiltonian, which causes trouble in proving gauge invariance 
\cite{ScharfNC}. In SQED based on DKP theory it was achieved, by an
effective approach, that this second order term does not contribute to
S-matrix, and thus it can be neglected when we construct the Feynman rules
for the theory \cite{Berestetskii, Pimentel 1}.

In this paper we will consider SQED-DKP in the framework of Epstein-Glaser
causal perturbative method. This method was formulated to give a
mathematical rigorous treatment of ultraviolet divergences in quantum field
theory. In this framework such divergences do not appear anywhere in the
calculations due to the correct splitting of causal distributions into its
advanced and retarded parts \cite{Epstein, Scharf}. Our goals are to obtain
a non-effective and mathematically well defined theory for SQED-DKP and
recover the results of the corresponding effective theory already obtained
in the usual perturbative approach. In addition, this work must be viewed as
an initial step in the attempts to rigorously establish the
renormalizability of the theory which is still an open question.

This paper is organized as follows. In section 2 we present the DKP theory.
In section 3 we review the main features of the Epstein-Glaser causal method
and present the basis to construct the second order S matrix. In section 4
we consider Compton scattering and address the question of how the
propagator of the effective theory emerges in the causal approach. In this
context the gauge invariance plays a crucial role. In section 5 we calculate
the scalar vacuum polarization tensor. In section 6 we calculate the
self-energy and, by using an Ward identity, also the vertex correction in
the limit of zero momentum transfer. In section 7 we make our concluding
remarks. The implications of gauge invariance are presented in the Appendix.

\section{Duffin-Kemmer-Petiau theory}

The free DKP theory is given by the Lagrangian \cite{Umezawa, Pimentel 2,
Berestetskii} 
\begin{equation}
{\cal L}=\frac{i}{2}\overline{\psi }\beta ^{\mu }\overleftrightarrow{%
\partial }_{\mu }\psi -m\overline{\psi }\psi ,  \label{lagr}
\end{equation}
\bigskip where $\psi $ is a multicomponent wave function, $\overline{\psi }%
=\psi ^{\dagger }\eta ^{0},$ and $\eta ^{0}=2\left( \beta ^{0}\right)
^{2}-1. $ $\beta ^{\mu }$ are a set of matrices $\left( \mu =0,1,2,3\right) $
satisfying the algebraic relations 
\begin{equation}
\beta ^{\mu }\beta ^{\nu }\beta ^{\rho }+\beta ^{\rho }\beta ^{\nu }\beta
^{\mu }=\beta ^{\mu }g^{\nu \rho }+\beta ^{\rho }g^{\mu \nu }.
\label{algebra}
\end{equation}
The equations of motion are then 
\begin{equation}
\left( i\beta ^{\mu }\partial _{\nu }-m\right) \psi =0\qquad \text{
and\qquad\ }\overline{\psi }\left( i\beta ^{\mu }\partial _{\nu }+m\right)
=0.  \label{dkp}
\end{equation}
It is known \cite{irreps} that the algebra (\ref{algebra}) has only three
irreducible representations, whose degrees are 1, 5 and 10. The first one is
trivial, having no physical content. The second and the third ones
correspond, respectively, to the scalar and vectorial representations. In
this work we shall restrict us to the scalar case.

Defining $q\!\!\!/$$=\beta ^{\mu }q_{\mu }$ and using relations (\ref
{algebra}) it can be shown that, for any four-vector $q$ the following
relation is satisfied 
\begin{equation}
q\!\!\!/^{2}(q\!\!\!/^{2}-q^{2})=0.  \label{vinculo}
\end{equation}
By using this relation and the plane wave solutions for the equations of
motion we can verify that $p^{2}=m^{2}.$

\bigskip As an ilustrative example of an explicit representation of (\ref
{algebra}) we can get 
\begin{eqnarray}
\beta ^{0} &=&\left( 
\begin{array}{ccccc}
0 & 0 & 0 & 0 & 1 \\ 
0 & 0 & 0 & 0 & 0 \\ 
0 & 0 & 0 & 0 & 0 \\ 
0 & 0 & 0 & 0 & 0 \\ 
1 & 0 & 0 & 0 & 0
\end{array}
\right) ;\quad \quad \beta ^{1}=\left( 
\begin{array}{ccccc}
0 & 0 & 0 & 0 & 0 \\ 
0 & 0 & 0 & 0 & 1 \\ 
0 & 0 & 0 & 0 & 0 \\ 
0 & 0 & 0 & 0 & 0 \\ 
0 & -1 & 0 & 0 & 0
\end{array}
\right) ;\quad  \nonumber \\
&&\quad  \label{irrep} \\
\quad \quad \beta ^{2} &=&\left( 
\begin{array}{ccccc}
0 & 0 & 0 & 0 & 0 \\ 
0 & 0 & 0 & 0 & 0 \\ 
0 & 0 & 0 & 0 & 1 \\ 
0 & 0 & 0 & 0 & 0 \\ 
0 & 0 & -1 & 0 & 0
\end{array}
\right) ;\quad \quad \beta ^{3}=\left( 
\begin{array}{ccccc}
0 & 0 & 0 & 0 & 0 \\ 
0 & 0 & 0 & 0 & 0 \\ 
0 & 0 & 0 & 0 & 0 \\ 
0 & 0 & 0 & 0 & 1 \\ 
0 & 0 & 0 & -1 & 0
\end{array}
\right) .  \nonumber
\end{eqnarray}
Using this specific representation we can construct explicitly projection
operators and show that not all components of the field $\psi $ are
independent. For a detailed explanation we refer to references \cite
{Umezawa, Pimentel 2}. Here we only quote the explicit form of $\psi $ in
this representation: 
\begin{equation}
\psi =\left( 
\begin{array}{c}
\frac{i}{\sqrt{m}}\partial _{0}\varphi \\ 
\frac{i}{\sqrt{m}}\partial _{1}\varphi \\ 
\frac{i}{\sqrt{m}}\partial _{2}\varphi \\ 
\frac{i}{\sqrt{m}}\partial _{3}\varphi \\ 
\sqrt{m}\varphi
\end{array}
\right) ,\quad (\square +m^{2})\varphi =0.  \label{wf}
\end{equation}
Thus we readily see that not all components of $\psi $ can be independent.
Actually, we can choose only two of these components as independent ones. We
shall make use of this explicit representation in the remainder of this work
only for ilustrating the arguments, being our results valid for any scalar
representation of relations (\ref{algebra}).

Now we apply the standard procedure of canonical quantization in the{\it \ }%
free Lagrangian (\ref{lagr}) and obtain \cite{Berestetskii} 
\[
\lbrack \psi _{a}^{-}(x),\overline{\psi ^{+}}_{b}(y)]=\frac{1}{i}%
S_{ab}^{+}(x-y)\text{ }\qquad \text{and\qquad }[\overline{\psi ^{-}}%
_{a}(x),\psi _{b}^{+}(y)]=-\frac{1}{i}S_{ba}^{-}(y-x)\text{ ,} 
\]
where $\psi ^{-}$ and $\psi ^{+}\ $ contains only annihilation and creation
operators, respectively and 
\begin{equation}
S_{ab}^{\pm }(x)=\frac{1}{m}[i\partial \!\!\!/(i\partial \!\!\!/+m)%
]_{ab}\triangle ^{\pm }(x),  \label{S}
\end{equation}
where 
\begin{equation}
\triangle ^{\pm }(x)=\frac{(\pm )i}{(2\pi )^{3}}\int d^{4}p\text{ }\delta
(p^{2}-m^{2})\theta (\pm p^{0})\text{e}^{-ip.x}  \label{delta}
\end{equation}
are the positive (negative) frequency parts of the Pauli-Jordan distribution 
\[
\triangle (x)=\triangle ^{+}(x)+\triangle ^{-}(x). 
\]
As it is well known, this later distribution has causal support \cite{Scharf}
and can be written as 
\begin{equation}
\triangle (x)=\triangle ^{\text{ret}}(x)-\triangle ^{\text{adv}}(x),
\label{PJ}
\end{equation}
where $\triangle ^{\text{ret}}(x)$ and $\triangle ^{\text{adv}}(x)$ have,
respectively, retarded and advanced supports with respect to the point $x.$
Analogously we define 
\begin{equation}
S(x)\triangleq S^{+}(x)+S^{-}(x).  \label{S+-}
\end{equation}
We can see directly that this distribution also has a causal support, since
it is a linear combination of derivatives of $\triangle (x),$ and the
differentiation of a causal distribution does not affect the causal property
of its support. Now, by (\ref{S}) and (\ref{PJ}), it is possible to write 
\begin{equation}
S(x)=S^{\text{ret}}(x)-S^{\text{adv}}(x)\text{ ,}  \label{Scausal}
\end{equation}
where 
\begin{equation}
S^{\text{ret}}(x)=\frac{1}{m}[i\partial \!\!\!/(i\partial \!\!\!/+m)%
]\triangle ^{\text{ret}}(x)\quad \text{and\quad }S^{\text{adv}}(x)=\frac{1}{m%
}[i\partial \!\!\!/(i\partial \!\!\!/+m)]\triangle ^{\text{adv}}(x)\text{ .}
\label{S=r-a}
\end{equation}
As we will see in the next section, the above splitting of $S(x)$ in
retarded and advanced parts is not the unique possible.

The interaction with electromagnetic field is introduced by the minimal
substitution $\partial _{\mu }\rightarrow \partial _{\mu }-ieA_{\mu }$ in
the Lagrangian (\ref{lagr}), which becomes 
\begin{eqnarray}
{\cal L} &=&{\cal L}_{\text{F}}+{\cal L}_{\text{I}}\text{ ;}  \label{Lint} \\
{\cal L}_{\text{F}} &=&\frac{i}{2}\overline{\psi }\beta ^{\mu }%
\overleftrightarrow{\partial _{\mu }}\psi -m\overline{\psi }\psi \text{ ;} \\
{\cal L}_{\text{I}} &=&e\overline{\psi }\beta ^{\mu }\psi A_{\mu }\text{ ,}
\label{inter}
\end{eqnarray}
where we have used $e>0$.

\section{The Epstein-Glaser causal approach}

\hspace{0.7cm}In the Epstein-Glaser's causal method \cite{Epstein} the $S$%
-matrix is constructed without any reference to Hamiltonian formalism, its
explicit form being obtained by making use of certain physical conditions --
with causality playing a major role. In this approach the S-matrix is viewed
as an operator-valued distribution given by the perturbative series

\begin{equation}
S(g) = 1 + \sum_{n=1}^\infty \frac{1}{n!} \int d^4x_1 \ldots d^4x_n T_n(x_1,
\ldots ,x_n)g(x_1) \ldots g(x_n)\; ,  \label{sma}
\end{equation}

\noindent where $g(x)$ is a c-number test function supposed to belong to the
Schwartz space, $g(x) \in {\cal S}({\rm {\bf R^4}})$. The symmetric $n$%
-point functions $T_n(X)$ ($X\equiv \{x_1,\ldots ,x_n\}$) are the basic
building blocks to be inductively constructed, from the knowledge of $T_1(x)$%
, by means of the requirements of causality

\begin{equation}
S(g_1+g_2)=S(g_1)S(g_2)\;, \hspace{1.0cm}{\rm if}\;\;\; {\rm supp}\; g_1 >%
{\rm supp}\; g_2\; ,  \label{ccau}
\end{equation}

\noindent and translational invariance 
\begin{equation}
U(a, {\bf 1})S(g)U(a, {\bf 1} )^{-1} = S\left( g(x-a)\right)\; .
\label{ctra}
\end{equation}

\noindent In the above equations $U(a,{\bf 1})$ is an usual representation
of the Poincar\'{e} group ${\cal P}_{+}^{\uparrow }$ in the Fock space and
the notation ${\rm supp}\;g_{1}>{\rm supp}\;g_{2}$ signify that all points
in the support of $g_{2}(x)$ occur at times previous than all points in the
support of $g_{1}(x)$.

In terms of the $n$-point functions $T_n$ these causality and translation
invariance conditions are given by

\begin{eqnarray}
T_n(x_1,\ldots ,x_n) =&& T_m(x_1, \ldots ,x_m)T_{n-m}(x_{m+1},\ldots ,x_n)\;
,  \label{caust} \\
&& \;\;\;\;{\rm if}\;\;\;\;\{ x_1, \ldots ,x_m \}> \{ x_{m+1},\ldots ,x_n
\}\; ,  \nonumber
\end{eqnarray}

\noindent and

\begin{equation}
U(a,{\bf 1})T_n(x_1,\ldots ,x_n) U(a,{\bf 1})^{-1} =T_n(x_1+a ,\ldots
,x_n+a) \; ,
\end{equation}

\noindent respectively.

Making use of these requirements we are able to construct the $T_{n}$, order
by order, from the explicit form of $T_{1}(x)$. It is known that, based on
general arguments such as {\it correspondence }\cite{Bogoliubov}, $T_{1}=i%
{\cal L}_{I}^{(1)},$ where ${\cal L}_{I}^{(1)}$ is the term of first order
in coupling constant in the interaction Lagrangian and is written in terms
of {\it free} fields \cite{ScharfNC}.

Now, let us sketch the inductive procedure (for a detailed account see \cite
{Scharf, Aste}). Suppose that all $T_{m}(X)$, with $m\leq n-1$ are known,
then one can define the distributions

\begin{eqnarray}
A_{n}^{^{\prime }}(x_{1},\ldots ,x_{n}) &=&\sum_{P_{2}}{\tilde{T}}%
_{n_{1}}(X)T_{n-n_{1}}(Y,x_{n}),  \nonumber \\
&&  \label{arli} \\
R_{n}^{^{\prime }}(x_{1},\ldots ,x_{n}) &=&\sum_{P_{2}}T_{n-n_{1}}(Y,x_{n}){%
\tilde{T}}_{n_{1}}(X),  \nonumber
\end{eqnarray}

\noindent where $P_2$ stands for all partitions $P_2 : \{ x_1, \ldots
,x_{n-1} \} = X \cup Y$, $X \neq \O $ into disjoint subsets with $\mid\!
X\!\mid =n_1$, $\mid\! Y\! \mid \leq n-2$. In (\ref{arli}) ${\tilde{T}}%
_{n}(X)$ refers to the $n$-point distributions corresponding to a series for
the $S^{-1}$-matrix analogous to (\ref{sma}) and can be obtained by formal
inversion of $S(g)$, giving

\begin{equation}
{\tilde{T}}_n(X) = \sum_{r=1}^n (-)^r \sum_{P_r} T_{n_1}(X_1)\ldots
T_{n_r}(X_r) ,  \label{tinv}
\end{equation}

\noindent where $P_r$ indicates all partitions of $X$ into $r$ disjoint
subsets: $X= X_1\cup \ldots \cup X_r$, $X_j \neq \O$, $|X_j|=n_j$. Of
course, since all $T_m(X)$, $m \leq n-1$ are given by the induction
hypothesis, also the $\tilde{T}_m(X)$ with $m \leq n-1$ are known.

If in (\ref{arli}) the sums are extended in order to include the empty set $%
X=\O $ we get

\begin{eqnarray}
A_{n}(x_{1},\ldots ,x_{n}) &=&\sum_{P_{2}^{0}}{\tilde{T}}%
_{n_{1}}(X)T_{n-n_{1}}(Y,x_{n})  \nonumber \\
&&  \nonumber \\
&=&A_{n}^{^{\prime }}(x_{1},\ldots ,x_{n})+T_{n}(x_{1},\ldots ,x_{n}), 
\nonumber \\
&&  \label{ar} \\
R_{n}(x_{1},\ldots ,x_{n}) &=&\sum_{P_{2}^{0}}T_{n-n_{1}}(Y,x_{n}){\tilde{T}}%
_{n_{1}}(X)  \nonumber \\
&&  \nonumber \\
&=&R_{n}^{^{\prime }}(x_{1},\ldots ,x_{n})+T_{n}(x_{1},\ldots ,x_{n}), 
\nonumber
\end{eqnarray}

\noindent where $P_{2}^{0}$ stands for all partitions $P_{2}^{0}:\{x_{1},%
\ldots ,x_{n-1}\}=X\cup Y$. A glance at equations (\ref{ar}) shows that $%
A_{n}$ and $R_{n}$ are not known because they contain the unknown $T_{n}$.
However, the distribution defined by

\begin{equation}
D_n (x_1, \ldots ,x_n) \equiv R_n -A_n=R^{^{\prime}}_n - A^{^{\prime}}_n ,
\label{d}
\end{equation}

\noindent is known.

Making use of causality it turns out that $R_n$ has retarded support and $%
A_n $ has advanced support, i.e.

\begin{equation}
{\rm supp} R_n (X) \subseteq \Gamma^{+}_{n-1}(x_n), \hspace{0.5cm} {\rm supp}
A_n (X) \subseteq \Gamma^{-}_{n-1}(x_n),  \label{arsup}
\end{equation}

\noindent with

\begin{eqnarray}
\Gamma^{\pm}_{n-1}(x) \equiv \{ (x_1, \ldots ,x_{n-1}) \mid x_j \in 
\overline{V}^{\pm} (x), \forall j= 1, \ldots ,n-1 \},  \nonumber \\
\\
\overline{V}^{\pm} (x) = \{ y \mid (y-x)^2 \geq 0 , \pm (y^0 - x^0 ) \geq 0
\}.  \nonumber
\end{eqnarray}

\noindent The distribution $D_n$ has causal support, ${\rm supp} D_n
\subseteq \Gamma^{+}_{n-1} \cup \Gamma^{-}_{n-1}$. In fact, a general proof
of the causal support of $D_n$ only exists for $n \geq 3$ -- for $n=2$ we
must verify this explicitly. Then, decomposing $D_n$ in advanced and
retarded distributions we obtain the $T_n$ distri{\-}bution by (\ref{ar}).

The operator-valued distributions which we shall have to split are of the
form

\begin{equation}
D_n(x_1,...,x_n) = \sum_k : \prod_j \overline{\psi}(x_j) d^k_n (x_1, \ldots
,x_n) \prod_l \psi(x_l) \prod_m A(x_m) :,
\end{equation}

\noindent where $\psi$, $\overline{\psi}$ are the free boson fields of DKP
theory and $A$ stands for the free gauge boson fields. In this expression $%
d_n^k$ are numerical tempered distributions, $d_n^k\in {\cal S}^{^{\prime}}(%
{\rm {\bf R^{4n}}})$, with causal support. Because of the translation
invariance, it is sufficient to put $x_n=0$ and consider

\begin{equation}
d(x) \equiv d_n^k (x_1,\ldots ,x_{n-1},0) \in {\cal S}^{^{\prime}}({\rm {\bf %
R^m}}), \;\;\; \; {\rm {\bf m}}=4{\rm {\bf n}}-4.
\end{equation}

The nontrivial step is the splitting of the numerical causal distribution $d$
in the advanced and retarded distributions $a$ and $r$, respectively. From
the fact that $\Gamma ^{+}(0)\cap \Gamma ^{-}(0)=\{0\}$ we can see that the
behaviour of $d(x)$ in $x\!=0$ (or, in momentum space, $p=\infty $) is
crucial in the splitting problem. With this in mind, a classification of the
distributions is given in which $d(x)\in {\cal S}^{^{\prime }}({\rm {\bf %
R^{m}}})$ is called singular of order $\omega $ if its Fourier transform $%
\hat{d}(p)$ has a quasi-asymptotics $\hat{d}_{0}(p)$ at $p=\infty $ \cite
{Scharf, Aste}

\begin{equation}
\lim_{\delta \rightarrow 0} \rho (\delta) \langle \hat{d}(\frac{p}{\delta }%
), \stackrel{\vee}{\phi}(p)\rangle = \langle \hat{d}_0, \stackrel{\vee}{\phi}
\rangle,  \label{d2}
\end{equation}

\noindent ($\stackrel{\vee}{\phi}$ is the inverse Fourier transform of $\phi 
$), with power-counting function $\rho (\delta )$ satisfying

\begin{equation}
\lim_{\delta \rightarrow 0} \frac{\rho ( c \delta )}{ \rho ( \delta )}=
c^{\omega},  \label{d3}
\end{equation}

\noindent for each $c>0$. Of course, there is an equivalent definition in
the coordinate space \cite{Scharf}, but, since the splitting is more easily
performed in the momentum space, this one is sufficient for our purposes.

Then, we have two distinct cases \cite{Epstein, Scharf}: $i)$ $\omega <0$ --
in this case the solution of the splitting problem is unique and the
retarded distribution can be defined by multiplication by step functions; $%
ii)$ $\omega \geq 0$ -- now the solution can be no longer obtained by
multiplying $d$ by step functions and, after a careful mathematical
treatment, it may be shown that the retarded distribution is given by the
``central splitting solution'' \cite{Scharf}

\begin{equation}
\hat{r} (p) = \frac{i}{2\pi}\int_{- \infty}^{+ \infty} dt \frac{\hat{d}(tp)}{%
(t-i0)^{\omega +1}(1-t+i0)}.  \label{cs}
\end{equation}

\noindent This solution has the very important feature that it preserves the
symmetries of the theory, in special Lorentz covariance and gauge
invariance. However, in contrast with the case ${\omega}<0$, the solution of
the splitting problem (\ref{cs}) is not the unique one and, in momentum
space, the general solution is given by

\begin{equation}
\tilde{r}(p) = \hat{r}(p) + \sum_{\mid a \mid = 0}^{\omega} C_a p^{a} ,
\label{gen}
\end{equation}

\noindent where the $C_a$ are constant coefficients which are not fixed by
the causal structure -- we need additional physical conditions in order to
determine them.

In expression (\ref{gen}) use is made of the {\it minimal distribution
splitting} condition which says that the singular order cannot be raised in
the splitting. This condition, crucial for a correct prediction of the
anomalous magnetic moment in QED$_{4}$ \cite{Scharf} and in the analysis of
the dynamic mass generation in (2+1) dimensions \cite{ScharPim, mpt}, will
also be useful here.

Now we apply the inductive steps above to construct the two-point
distributions for SQED-DKP theory. Then, the one-point distribution for the
DKP field interacting with electromagnetic field is given formally by $i$
times the interaction term (\ref{inter}) in the Lagrangian (\ref{Lint}), 
\begin{equation}
T_{1}(x)=ie:\overline{\psi }(x)\beta ^{\mu }\psi (x):A_{\mu }(x)=-\widetilde{%
T}_{1}(x)\text{ ,}  \label{interact}
\end{equation}
where\ all fields entering in this expression are {\it free} fields and $e$
is the {\it physical} charge. The normal ordering is necessary in order to
have a well defined expression for the product of field operators at the
same point.

To go from $n=1$ to $n=2$ we take (\ref{interact}) and construct the
distributions $A_{2}^{^{\prime }}(x_{1},x_{2})$ and $R_{2}^{\prime
}(x_{1},x_{2})$ from expressions (\ref{ar}). So, we have 
\begin{eqnarray}
A_{2}^{^{\prime }}(x_{1},x_{2}) &=&e^{2}:\overline{\psi }(x_{1})\beta ^{\mu
}\psi (x_{1})::\overline{\psi }(x_{2})\beta ^{\nu }\psi (x_{2}):A_{\mu
}(x_{1})A_{\nu }(x_{2})\text{;}  \label{A2} \\
R_{2}^{^{\prime }}(x_{1},x_{2}) &=&e^{2}\text{ }:\overline{\psi }%
(x_{2})\beta ^{\nu }\psi (x_{2})::\overline{\psi }(x_{1})\beta ^{\mu }\psi
(x_{1}):A_{\mu }(x_{1})A_{\nu }(x_{2})\text{ ,}  \label{R2}
\end{eqnarray}
and then 
\begin{eqnarray}
D_{2}(x_{1},x_{2}) &=&e^{2}\{:\overline{\psi }(x_{2})\beta ^{\nu }\psi
(x_{2})::\overline{\psi }(x_{1})\beta ^{\mu }\psi (x_{1}): \\
&&-:\overline{\psi }(x_{1})\beta ^{\mu }\psi (x_{1})::\overline{\psi }%
(x_{2})\beta ^{\nu }\psi (x_{2}):\}A_{\mu }(x_{1})A_{\nu }(x_{2})\text{.}
\label{d=r-a}
\end{eqnarray}
$\!$ By using Wick's Theorem this expression can be written as a sum of
terms, each of them consisting of a product of field operators and Wick
contractions \cite{Scharf}. In this work we shall be concerned to the terms
corresponding to Compton scattering, vacuum polarization and self-energy.
From these we will determine the vertex correction via{\it \ }an Ward
identity. These terms are given by 
\begin{eqnarray}
D_{2}^{\text{Com}}(x_{1},x_{2}) &=&e^{2}\beta _{ab}^{\mu }\beta _{cd}^{\nu }
\nonumber \\
&&\hspace{-2.5cm}\times \{:\overline{\psi }_{a}(x_{1})\psi
_{d}(x_{2})::A_{\mu }(x_{1})A_{\nu }(x_{2}):[\overbrace{\overline{\psi }%
_{c}(x_{2})\psi _{b}}(x_{1})-\overbrace{\psi _{b}(x_{1})\overline{\psi }_{c}}%
(x_{2})]  \nonumber \\
&&\hspace{-2.5cm}+:\psi _{b}(x_{1})\overline{\psi }_{c}(x_{2})::A_{\mu
}(x_{1})A_{\nu }(x_{2}):[\overbrace{\psi _{d}(x_{2})\overline{\psi }_{a}}%
(x_{1})-\overbrace{\overline{\psi }_{a}(x_{1})\psi _{d}}(x_{2})]\}\text{ ;}
\label{Compton} \\
&&  \nonumber \\
D_{2}^{\text{Vac}}(x_{1},x_{2}) &=&e^{2}\beta _{ab}^{\mu }\beta _{cd}^{\nu
}:A_{\mu }(x_{1})A_{\nu }(x_{2}):  \nonumber \\
&&\times \lbrack \overbrace{\overline{\psi }_{c}(x_{2})\psi _{b}}(x_{1})%
\overbrace{\psi _{d}(x_{2})\overline{\psi }_{a}}(x_{1})-\overbrace{\overline{%
\psi }_{a}(x_{1})\psi _{d}}(x_{2})]\text{ ,}  \label{Polva} \\
&&  \nonumber \\
D_{2}^{\text{self}}(x_{1},x_{2}) &=&e^{2}\beta _{ab}^{\mu }\beta _{cd}^{\nu }
\nonumber \\
&&\times \{ :\overline{\psi }_{a}(x_{1})\psi _{d}(x_{2}):[\overbrace{%
\overline{\psi }_{c}(x_{2})\psi _{b}}(x_{1})\overbrace{A_{\nu }(x_{2})A_{\mu
}}(x_{1})  \nonumber \\
&&-\overbrace{\psi _{b}(x_{1})\overline{\psi }_{c}}(x_{2})\overbrace{A_{\mu
}(x_{1})A_{\nu }}(x_{2})]  \nonumber \\
&&+ :\psi _{b}(x_{1})\overline{\psi }_{c}(x_{2}):[\overbrace{\psi _{d}(x_{2})%
\overline{\psi }_{a}}(x_{1})\overbrace{A_{\nu }(x_{2})A_{\mu }}(x_{1}) 
\nonumber \\
&&-\overbrace{\overline{\psi }_{a}(x_{1})\psi _{d}}(x_{2})\overbrace{A_{\mu
}(x_{1})A_{\nu }}(x_{2})]\}\text{ ,}  \label{self}
\end{eqnarray}
where the Wick contractions are defined as 
\begin{eqnarray}
\overbrace{\psi _{a}(x)\overline{\psi }_{b}}(y) &\triangleq &[\psi
_{a}^{-}(x),\overline{\psi }_{b}^{+}(y)]=\frac{1}{i}S_{ab}^{+}(x-y)\text{ ;}
\label{c1} \\
\overbrace{\overline{\psi }_{a}(x)\psi _{b}}(y) &\triangleq &[\overline{\psi
_{a}^{-}}(x),\psi _{b}^{+}(y)]=-\frac{1}{i}S_{ba}^{-}(y-x)\text{ ;}
\label{c2} \\
\overbrace{A_{\mu }(x)A_{\nu }}(y) &\triangleq &\text{ }[A_{\mu
}^{-}(x),A_{\nu }^{+}(y]\text{ \ \ }=ig_{\mu \nu }D_{0}^{+}(x-y)\text{ ,}
\label{c3}
\end{eqnarray}
and $S_{ab}^{+}(x)$ and $S_{ba}^{-}(x)$ are given by (\ref{S}). $%
D_{0}^{+}(x) $ is the positive frequency part of the zero mass Pauli-Jordan
distribution, $D_{0}^{+}(x)=\frac{i}{(2\pi )^{3}}\int d^{4}p$ $\delta
(p^{2})\theta (p^{0}) $e$^{-ip.x}$.

\section{The Compton scattering}

We denote respectively by $D_{2}^{\text{I}}(x_{1},x_{2})$ and $D_{2}^{\text{%
II}}(x_{1},x_{2})$ the first and second terms inside curl brackets in (\ref
{Compton}). By using (\ref{c1}) and \ref{c2}) we have 
\begin{eqnarray}
D_{2}^{\text{I}}(x_{1},x_{2}) &=&ie^{2}\beta _{ab}^{\mu }\beta _{cd}^{\nu }:%
\overline{\psi }_{a}(x_{1})\psi _{d}(x_{2})::A_{\mu }(x_{1})A_{\nu }(x_{2}):
\nonumber \\
&&\times \{S_{bc}^{-}(x_{1}-x_{2})+S_{bc}^{+}(x_{1}-x_{2})\}  \label{a} \\
&=&ie^{2}\beta _{ab}^{\mu }\beta _{cd}^{\nu }:\overline{\psi }%
_{a}(x_{1})\psi _{d}(x_{2})::A_{\mu }(x_{1})A_{\nu
}(x_{2}):S_{bc}(x_{1}-x_{2})  \label{D2}
\end{eqnarray}
where we used the definition (\ref{S+-}). The first term inside curl
brackets in (\ref{a}) cames from $R_{2}^{\prime \text{I}}(x_{1},x_{2}),$
whereas the second cames from $A_{2}^{\prime \text{I}}(x_{1},x_{2})$.
Similarly, 
\begin{eqnarray}
D^{\text{II}}(x_{1},x_{2}) &=&ie^{2}\beta _{ab}^{\mu }\beta _{cd}^{\nu
}:\psi _{b}(x_{1})\overline{\psi }_{c}(x_{2})::A_{\mu }(x_{1})A_{\nu
}(x_{2}):  \nonumber \\
&&\times \{-S_{da}^{+}(x_{2}-x_{1})-S_{da}^{-}(x_{2}-x_{1})\}  \label{b} \\
&=&ie^{2}\beta _{ab}^{\mu }\beta _{cd}^{\nu }:\psi _{b}(x_{1})\overline{\psi 
}_{c}(x_{2})::A_{\mu }(x_{1})A_{\nu }(x_{2}):  \nonumber \\
&&\times \{-S_{da}(x_{2}-x_{1})\}\text{,}  \label{D4}
\end{eqnarray}
where again the first term inside curl brackets in (\ref{b}) cames from $%
R_{2}^{\prime \text{II}}(x_{1},x_{2})$ and so on. As we said in Section 2,
the distribution $S(x)$ has causal support. Then, we have verified
explicitly that the distributions (\ref{D2}) and (\ref{D4}) have causal
support too.

Since $S(x)$ itself is the numerical distribution we must split into
retarded and advanced parts in equations (\ref{D2}) and (\ref{D4}), a
splitting solution is trivially obtained from (\ref{Scausal}). But, since $%
S(x)$ has singular order $\omega =0$ --as we can verify by (\ref{d2})and (%
\ref{d3}) -- this splitting is not unique. So, accordingly with (\ref{gen}),
the general solution for the retarded distribution in configuration space is 
\[
\tilde{r}\left( x\right) =S^{\text{ret}}(x)+C\delta (x)\text{,} 
\]
where $C$ is an arbitrary constant. Now we construct the numerical
distribution $t^{\text{I}}(x_{1},x_{2})$ from (\ref{ar}), 
\begin{equation}
t^{\text{I}}(x_{1},x_{2})=\tilde{r}(x_{1},x_{2})-r^{\prime }(x_{1},x_{2})%
\text{ ,}  \label{jj}
\end{equation}
where $r^{\prime }(x_{1},x_{2})$ is the numerical distribution associated
with $R_{2}^{\prime \text{I}}(x_{1},x_{2})$, which is obtained from the
first term inside curl brackets in (\ref{a}). Then, 
\begin{eqnarray}
t^{\text{I}}(x_{1},x_{2}) &=&S^{\text{ret}%
}(x_{1}-x_{2})-S^{-}(x_{1}-x_{2})+C\delta (x_{1}-x_{2})  \nonumber \\
&=&-S^{\text{F}}(x_{1}-x_{2})+C\delta (x_{1}-x_{2})\text{ ,}  \label{A}
\end{eqnarray}
where we have defined 
\begin{equation}
-S^{\text{F}}(x)\triangleq S^{\text{ret}}(x)-S^{-}(x)=S^{\text{adv}%
}(x)+S^{+}(x)\text{ }=-\frac{1}{m}i\partial \!\!\!/(i\partial
\!\!\!/+m)\bigtriangleup ^{\text{F}}(x)\text{ ,}  \label{A'}
\end{equation}
where \ $\bigtriangleup ^{\text{F}}(x)$ is the usual Feynman scalar
propagator.

In a similar way we find (note that the reference point for splitting is $%
x_{2}$) 
\begin{eqnarray}
t^{\text{II}}(x_{1},x_{2}) &=&S^{\text{adv}%
}(x_{2}-x_{1})+S^{+}(x_{2}-x_{1})+C^{\prime }\delta (x_{2}-x_{1})  \nonumber
\\
&=&-S^{\text{F}}(x_{2}-x_{1})+C^{\prime }\delta (x_{2}-x_{1})\text{ .}
\label{B}
\end{eqnarray}
The constants $C$ and $C^{\prime }$ will be determined later by the
requirements of charge conjugation and gauge invariance.

Now the complete two-point distribution for Compton scattering is given by 
\begin{equation}
T_{2}^{\text{Compton}}(x_{1},x_{2})=T_{2}^{\text{I}}(x_{1},x_{2})+T_{2}^{%
\text{II}}(x_{1},x_{2}),  \label{total}
\end{equation}
where 
\[
T_{2}^{\text{I}}(x_{1},x_{2})=ie^{2}:\overline{\psi }(x_{1})\beta ^{\mu }t^{%
\text{I}}(x_{1},x_{2})\beta ^{\nu }\psi (x_{2})::A_{\mu }(x_{1})A_{\nu
}(x_{2}):
\]
and 
\[
T_{2}^{\text{II}}(x_{1},x_{2})=ie^{2}:\overline{\psi }(x_{2})\beta ^{\nu }t^{%
\text{II}}(x_{1},x_{2})\beta ^{\mu }\psi (x_{1})::A_{\mu }(x_{1})A_{\nu
}(x_{2}):.
\]

It is very simple to require charge conjugation invariance of the theory 
\cite{Berestetskii, Scharf}. Here we only quote the condition that arise
from this invariance, namely 
\[
C=C^{\prime }. 
\]
Thus we can immediately see that 
\[
T_{2}^{\text{I}}(x_{1},x_{2})=T_{2}^{\text{II}}(x_{2},x_{1}), 
\]
what implies that $T_{2}^{\text{Compton}}(x_{1},x_{2})$ is symmetric in its
arguments, as it would be.

In the Appendix we calculate the remaining constant $C$ by requiring gauge
invariance of the theory. There we find (see eq. (\ref{AC})) 
\[
C=\frac{I}{m}\text{ ,} 
\]
where $\ I$ is the $5\times 5$ identity matrix. Turning this result into (%
\ref{A}) and (\ref{B}) we obtain 
\begin{eqnarray*}
t^{\text{I}}(x_{1},x_{2}) &=&-\{S^{\text{F}}(x_{1}-x_{2})-\frac{\hat{I}}{m}%
\delta (x_{1}-x_{2})\}\text{ } \\
&=&t^{\text{II}}(x_{2},x_{1})\text{ .}
\end{eqnarray*}
We denote the distribution inside the curl brackets in the above expressions
as 
\begin{equation}
T^{\text{c}}(x)\triangleq S^{\text{F}}(x)-\frac{\hat{I}}{m}\delta (x)\text{ .%
}  \label{ep}
\end{equation}
It is direct to see that this distribution is the Green function for the DKP
equation, i. e., 
\begin{equation}
(i\partial \!\!\!/-m)T^{\text{c}}(x)=\delta (x)\text{ .}  \label{GF}
\end{equation}
So, in this sense the distribution (\ref{ep}) is the \ correct \ \
``propagator'' for the DKP scalar particle, that is the same as the \
``effective'' propagator of \ references \cite{Berestetskii, Pimentel 1}.
For later use, we write this propagator in the momentum space 
\begin{equation}
\widehat{T}^{\text{c}}(p)=\frac{1}{(2\pi )^{2}m}\left[ \frac{%
p\!\!\!/(p\!\!\!/+m)}{p^{2}-m^{2}+i0}-1\right] \text{ .}  \label{prop}
\end{equation}

With the above results we write finally the two-point distribution for the
Compton scattering: 
\begin{eqnarray}
T_{2}^{\text{Compton}}(x_{1},x_{2}) &=&-ie^{2}\{:\overline{\psi }%
(x_{1})\beta ^{\mu }T^{\text{c}}(x_{1}-x_{2})\beta ^{\nu }\psi (x_{2}): 
\nonumber \\
&&+:\overline{\psi }(x_{2})\beta ^{\nu }T^{\text{c}}(x_{2}-x_{1})\beta ^{\mu
}\psi (x_{1}):\}:A_{\mu }(x_{1})A_{\nu }(x_{2}):.  \label{Compfinal}
\end{eqnarray}

\section{The vacuum polarization}

Now we consider the term (\ref{Polva}), associated with scalar vacuum
polarization. After substituting the explicit form of Wick contractions,
this term is given by 
\begin{eqnarray}
D_{2}^{\text{Vac}}(x_{1},x_{2}) &=&e^{2}:A_{\mu }(x_{1})A_{\nu }(x_{2}): 
\nonumber \\
&&\hspace{-2cm}\times \text{Tr}\{\beta ^{\mu }S^{-}(x_{1}-x_{2})\beta ^{\nu
}S^{+}(x_{2}-x_{1})-\beta ^{\nu }S^{-}(x_{2}-x_{1})\beta ^{\mu
}S^{+}(x_{1}-x_{2})\}\text{ ,}  \nonumber \\
&&  \label{D6}
\end{eqnarray}
where the first term inside curl brackets cames from $R^{\prime
}(x_{1}-x_{2})$ and the second from $A^{\prime }(x_{1}-x_{2})$. Making $%
y=x_{1}-x_{2}$ and defining 
\[
P^{\mu \nu }(y)\triangleq -e^{2}\text{Tr}\{\beta ^{\mu }S^{+}(y)\beta ^{\nu
}S^{-}(-y)\}\text{ ,} 
\]
we write (\ref{D6}) in the form 
\begin{equation}
D_{2}^{\text{Vac}}(x_{1},x_{2})=\{P^{\mu \nu }(y)-P^{\nu \mu }(-y)\}:A_{\mu
}(x_{1})A_{\nu }(x_{2}):\text{ .}  \label{D6P}
\end{equation}
This distribution has a causal support, as required. To see this we use the
fact that $S(x)$ has a causal support and write the term inside curl
brackets in the above expression as 
\[
\beta ^{\mu }[S(y)\beta ^{\nu }S^{+}(-y)-S^{+}(-y)\beta ^{\nu }S(-y)]\text{
\ .} 
\]

Before splitting the causal distribution above, we consider the Fourier
transform of $P^{\mu \nu }(y)$, 
\begin{eqnarray*}
\hat{P}^{\mu \nu }(k) &=&\frac{1}{(2\pi )^{2}}\int dyP^{\mu \nu }(y)\text{e}%
^{ik.y} \\
&=&\frac{-e^{2}}{(2\pi )^{2}}\text{Tr}\int dy\beta ^{\mu }S^{+}(y)\beta
^{\nu }S^{-}(-y)\text{e}^{ik.y}\text{ .}
\end{eqnarray*}
Substituting into this expression the explicit forms of $S^{+}(y)$ and $%
S^{-}(-y)$ given by (\ref{S}) and (\ref{delta}), we have 
\begin{eqnarray}
\hat{P}^{\mu \nu }(k) &=&\frac{-e^{2}}{(2\pi )^{4}m^{2}}\int dp\text{Tr}%
\{\beta ^{\mu }p\!\!\! /(p\!\!\!/+m)\beta ^{\nu }(\NEG{k}-p\!\!\!/)(\NEG%
{k}-p\!\!\!/-m)\}  \nonumber \\
&&  \label{PMN} \\
&&\times \theta (p^{0})\delta (p^{2}-m^{2})\theta (k^{0}-p^{0})\delta
\lbrack (k-p)^{2}-m^{2}].  \nonumber
\end{eqnarray}
To compute the trace inside this integral we need the following trace
properties \cite{trace}: 
\begin{eqnarray*}
\text{Tr}\{\beta ^{\mu _{1}}\beta ^{\mu _{2}}...\beta ^{\mu _{2n-1}}\} &=&0%
\text{ ;} \\
\text{Tr}\{\beta ^{\mu _{1}}\beta ^{\mu _{2}}...\beta ^{\mu _{2n}}\}
&=&g^{\mu _{1}\mu _{2}}g^{\mu _{3}\mu _{4}}...g^{\mu _{2n-1}\mu
_{2n}}+g^{\mu _{2}\mu _{3}}g^{\mu _{4}\mu _{5}}...g^{\mu _{2n}\mu _{1}}\text{
,}
\end{eqnarray*}
with $n=1,2,3,...$. So, we obtain 
\[
\text{Tr}\{\beta ^{\mu }p\!\!\!/(p\!\!\!/+m)\beta ^{\nu }(\NEG{k}-p\!\!\!/)(%
\NEG{k}-p\!\!\!/-m)\}=m^{2}[4p^{\mu }p^{\nu }-2(k^{\mu }p^{\nu }+p^{\mu
}k^{\nu })+k^{\mu }k^{\nu }]\text{ .} 
\]
When substituting this result into (\ref{PMN}), and taking into account the
distribution $\delta \lbrack (k-p)^{2}-m^{2}],$ we observe that $\hat{P}%
^{\mu \nu }(k)$ is a second rank symmetric tensor which satisfies 
\[
k_{\mu }\hat{P}^{\mu \nu }(k)=0\text{ ,} 
\]
what means that the vacuum polarization term (\ref{D6P}) is gauge invariant.
Using this fact, we can write, as usual, 
\[
\hat{P}^{\mu \nu }(k)=(k^{\mu }k^{\nu }-k^{2}g^{\mu \nu })B(k^{2})\text{ ,} 
\]
where 
\[
B(k^{2})=\frac{-1}{3k^{2}}\hat{P}_{\mu }^{\mu }(k)\text{ .} 
\]
The computation of $\hat{P}_{\mu }^{\mu }(k)$ is straightforward and the
final result for $\hat{P}^{\mu \nu }(k)$ is 
\begin{equation}
\hat{P}^{\mu \nu }(k)=\frac{-e^{2}}{(2\pi )^{4}}\left( \frac{k^{\mu }k^{\nu }%
}{k^{2}}-g^{\mu \nu }\right) \frac{\pi k^{2}}{6}\left( 1-\frac{4m^{2}}{k^{2}}%
\right) ^{\frac{3}{2}}\theta \left( k^{2}-4m^{2}\right) \theta (k^{0})\text{
.}  \label{Pmnf}
\end{equation}

Now, in momentum space the numerical distribution associated with (\ref{D6P}%
), which we must split, is given by 
\begin{equation}
\hat{P}^{\mu \nu }(k)-\hat{P}^{\mu \nu }(-k)=\frac{-e^{2}}{(2\pi )^{4}}%
\left( \frac{k^{\mu }k^{\nu }}{k^{2}}-g^{\mu \nu }\right) \hat{d}(k)\text{ ,}
\label{rr}
\end{equation}
where 
\begin{equation}
\hat{d}(k)=\frac{\pi k^{2}}{6}\left( 1-\frac{4m^{2}}{k^{2}}\right) ^{\frac{3%
}{2}}\theta \left( k^{2}-4m^{2}\right) \text{sgn}(k^{0})\text{ .}  \label{dk}
\end{equation}
As the tensor character of (\ref{rr}) does not matter in its splitting
procedure, the problem reduces to the splitting of $\ \hat{d}(k)$. To do
this, we first determine the singular order $\omega $ of this distribution
by using (\ref{d2}) and (\ref{d3}). So, we find $\omega =2.$ To simplify the
calculations we make use of the fact that $k$ in (\ref{dk}) is timelike. So,
we can choose a Lorentz frame such that $k=(k^{0},\vec{0}).$ In this case
the central splitting formula (\ref{cs}) can be written as \cite{Scharf}: 
\[
\hat{r}(k^{0})=\frac{i}{2\pi }\left( k^{0}\right) ^{3}\int_{-\infty
}^{+\infty }dp^{0}\frac{\hat{d}(p^{0})}{(p^{0}-i0)^{3}(k^{0}-p^{0}+i0)}.
\]
Substituting (\ref{dk}) into this integral and making the substitution $%
s=(p^{0})^{2},$ we obtain 
\begin{eqnarray*}
\hat{r}(k^{0}) &=&\frac{i}{12}\left( k^{0}\right) ^{4}\left\{ \text{P.V.}%
\int\limits_{4m^{2}}^{\infty }ds\frac{1}{s\left[ \left( k^{0}\right) ^{2}-s%
\right] }\left( 1-\frac{4m^{2}}{s}\right) ^{\frac{3}{2}}\right.  \\
&&\left. -i\pi \text{sgn}(k^{0})\theta \lbrack \left( k^{0}\right)
^{2}-4m^{2}]\frac{1}{\left( k^{0}\right) ^{2}}\left( 1-\frac{4m^{2}}{\left(
k^{0}\right) ^{2}}\right) ^{\frac{3}{2}}\right\} .
\end{eqnarray*}
The distribution $\hat{r}^{\prime }(k^{0})$ comes from the term $-\hat{P}%
^{\mu \nu }(-k)$ in (\ref{D6P}). So, 
\[
\hat{r}^{\prime }(k^{0})=-\frac{\pi \left( k^{0}\right) ^{2}}{6}\left( 1-%
\frac{4m^{2}}{\left( k^{0}\right) ^{2}}\right) ^{\frac{3}{2}}\theta \left[
\left( k^{0}\right) ^{2}-4m^{2}\right] \theta (-k^{0}).
\]
Now the two-point distribution is given, in an arbitrary Lorentz frame, as 
\begin{eqnarray*}
\hat{t}(k) &=&\hat{r}(k)-\hat{r}^{\prime }(k) \\
&=&\frac{i}{12}k^{4}\int\limits_{4m^{2}}^{\infty }ds\frac{1}{s(k^{2}-s+i0)}%
\left( 1-\frac{4m^{2}}{s}\right) ^{\frac{3}{2}}.
\end{eqnarray*}
Finally, the two-point distribution for the vacuum polarization in
configuration space is given by 
\[
T_{2}^{\text{Vac}}(x_{1},x_{2})=-i:A_{\mu }(x_{1})\Pi ^{\mu \nu
}(x_{1}-x_{2})A_{\nu }(x_{2}):\text{,}
\]
where 
\begin{eqnarray*}
\hat{\Pi}^{\mu \nu }(k) &=&\frac{-ie^{2}}{(2\pi )^{4}}\left( \frac{k^{\mu
}k^{\nu }}{k^{2}}-g^{\mu \nu }\right) \hat{t}(k) \\
&=&\frac{1}{(2\pi )^{4}}\left( \frac{k^{\mu }k^{\nu }}{k^{2}}-g^{\mu \nu
}\right) \hat{\Pi}(k)
\end{eqnarray*}
and 
\begin{equation}
\hat{\Pi}(k)=\frac{e^{2}}{12}k^{4}\int\limits_{4m^{2}}^{\infty }ds\frac{1}{%
s(k^{2}-s+i0)}\left( 1-\frac{4m^{2}}{s}\right) ^{\frac{3}{2}}\text{ .}
\label{res}
\end{equation}

For $0<k^{2}<4m^{2},$ the result of this integral can be writen as 
\[
\hat{\Pi}(k)=\frac{-e^{2}}{2}k^{2}\left[ \frac{4m^{2}-k^{2}}{3k^{2}}\left(
\theta \text{cot}\theta -1\right) +\frac{1}{9}\right] \text{ ,} 
\]
where sin$^{2}\theta =\frac{k^{2}}{4m^{2}}$ and $\theta \in (0,\frac{\pi }{2}%
).$

For spacelike $k$ we get 
\[
\hat{\Pi}(k)=\frac{e^{2}m^{2}}{12\lambda (1-\lambda )}\left[ \left(
1+\lambda \right) ^{3}\text{ln}\lambda +\frac{8}{3}(1-\lambda ^{3})\right] , 
\]
where $k^{2}=-\frac{(1-\lambda )^{2}}{\lambda }m^{2}$, and $0<\lambda <1.$
This result can be analytically extended for $k^{2}>4m^{2}$ by making $%
-1<\lambda <0$ and taking ln$\lambda =$ln$|\lambda |-i\pi $.

These results are not the most general solutions for the splitting problem,
as we had to split a distribution with singular order $\omega =2.$ The most
general solution $\tilde{\Pi}(k)$ is given by 
\[
\tilde{\Pi}(k)=\hat{\Pi}(k)+C_{0}+C_{\mu }k^{\mu }+C_{2}k^{2}, 
\]
where the normalization constants $C_{0}$, $C_{\mu }$ and $C_{2}$ are not
determined by causality. They are determined from the requirements of a zero
mass for the gauge field, parity invariance and the identification of $m$
with the physical observable mass. The procedure is the standard one and we
simply quote the results: $C_{0}=C_{\mu }=C_{2}=0.$

\section{The self-energy and vertex corrections}

The self-energy of DKP particle is given by (\ref{self}). After substituting
the Wick contractions we obtain the following two terms: 
\begin{eqnarray}
D_{\text{I}}^{\text{self}}\left( x_{1},x_{2}\right) &=&-e^{2}:\overline{\psi 
}(x_{1})\beta ^{\mu }\left[ S^{-}\left( x_{1}-x_{2}\right) D_{0}^{+}\left(
x_{2}-x_{1}\right) \right.  \nonumber \\
&&\left. +S^{+}\left( x_{1}-x_{2}\right) D_{0}^{+}\left( x_{1}-x_{2}\right)
\beta _{\mu }\psi (x_{2}):\right] \text{ ;}  \label{DIII} \\
D_{\text{II}}^{\text{self}}\left( x_{1},x_{2}\right) &=&e^{2}:\overline{\psi 
}(x_{2})\beta ^{\mu }\left[ S^{+}\left( x_{2}-x_{1}\right) D_{0}^{+}\left(
x_{2}-x_{1}\right) \right.  \nonumber \\
&&\left. +S^{-}\left( x_{2}-x_{1}\right) D_{0}^{+}\left( x_{1}-x_{2}\right)
\beta _{\mu }\psi (x_{1}):\right] \text{ .}  \label{DV}
\end{eqnarray}
Therefore we note that 
\begin{equation}
D_{\text{II}}^{\text{self}}\left( x_{1},x_{2}\right) =-D_{\text{I}}^{\text{%
self}}\left( x_{2},x_{1}\right) \text{ .}  \label{sim}
\end{equation}

Let us first consider $D_{\text{I}}^{\text{self}}\left( x_{1},x_{2}\right) $%
. From it we can determine the two-point distribution $T_{\text{I}}^{\text{%
self}}\left( x_{1},x_{2}\right) .$ From 
\begin{eqnarray*}
D_{\text{I}}^{\text{self}}\left( x_{1},x_{2}\right)  &=&R_{\text{I}}^{\prime
}\left( x_{1},x_{2}\right) -A_{\text{I}}^{\prime }\left( x_{1},x_{2}\right) 
\\
&=&R_{\text{I}}\left( x_{1},x_{2}\right) -A_{\text{I}}\left(
x_{1},x_{2}\right) \text{ }
\end{eqnarray*}
and 
\begin{eqnarray*}
D_{\text{II}}^{\text{self}}\left( x_{1},x_{2}\right)  &=&R_{\text{II}%
}^{\prime }\left( x_{1},x_{2}\right) -A_{\text{II}}^{\prime }\left(
x_{1},x_{2}\right)  \\
&=&R_{\text{II}}\left( x_{1},x_{2}\right) -A_{\text{II}}\left(
x_{1},x_{2}\right) \text{ ,}
\end{eqnarray*}
and using (\ref{sim}) we conclude that 
\begin{eqnarray*}
R_{\text{II}}^{\prime }\left( x_{1},x_{2}\right)  &=&A_{\text{I}}^{\prime
}\left( x_{2},x_{1}\right) \text{ ;} \\
A_{\text{II}}^{\prime }\left( x_{1},x_{2}\right)  &=&R_{\text{I}}^{\prime
}\left( x_{2},x_{1}\right) \text{ ;} \\
R_{\text{II}}\left( x_{1},x_{2}\right)  &=&A_{\text{I}}\left(
x_{2},x_{1}\right) \text{ ;} \\
A_{\text{II}}\left( x_{1},x_{2}\right)  &=&R_{\text{I}}\left(
x_{2},x_{1}\right) \text{ .}
\end{eqnarray*}

So, we have 
\begin{equation}
T_{\text{I}}^{\text{self}}\left( x_{1},x_{2}\right) =T_{\text{II}}^{\text{%
self}}\left( x_{2},x_{1}\right)   \label{sim2}
\end{equation}
and conclude that the two-point distribution for the self energy $T^{\text{%
self}}\left( x_{1},x_{2}\right) =T_{\text{I}}^{\text{self}}\left(
x_{1},x_{2}\right) +T_{\text{II}}^{\text{self}}\left( x_{1},x_{2}\right) $
is symmetric, as it would be. Moreover it suffices to compute one of the
distributions, say $T_{\text{I}}^{\text{self}}\left( x_{1},x_{2}\right) $,
and the other will be given by (\ref{sim2}).

Returning to $D_{\text{I}}^{\text{self}}\left( x_{1},x_{2}\right) $ in (\ref
{DIII}), we must first verify the causal property of the support of this
distribution. The term into brackets can be writen as 
\[
S\left( x_{1}-x_{2}\right) D_{0}^{+}\left( x_{2}-x_{1}\right) +S^{+}\left(
x_{1}-x_{2}\right) D_{0}\left( x_{1}-x_{2}\right) \text{ } 
\]
and,\hfill as \hfill $S\left( x_{1}-x_{2}\right) $ \hfill and \hfill $D_{0}\left( x_{1}-x_{2}\right) $ \hfill
are \hfill causal \hfill distributions, \hfill so \hfill it \hfill is \hfill also\hfill \\ $D_{\text{I}}^{\text{self}}\left(
x_{1},x_{2}\right) $.

The numerical distribution we have to split is 
\begin{equation}
d_{\text{I}}(y)=-e^{2}\beta ^{\mu }\left[
S^{-}(y)D_{0}^{+}(-y)+S^{+}(y)D_{0}^{+}(y)\right] \beta _{\mu }\text{ .}
\label{6}
\end{equation}
We denote the two terms into brackets in this expression by 
\begin{equation}
d_{-}(y)=S^{-}(y)D^{+}(-y)=-S^{-}(y)D^{-}(y)  \label{d-}
\end{equation}
and 
\begin{equation}
d_{+}(y)=S^{+}(y)D^{+}(y)\text{ .}  \label{d+}
\end{equation}

Taking the Fourier transform of $d_{-}$ and using the explicit form of
Pauli-Jordan distributions we obtain 
\begin{equation}
\widehat{d}_{-}(p)=\frac{1}{(2\pi )^{4}m}\int dq\text{ }\slash\!\!\!q\left( %
\slash\!\!\!q+m\right) \theta \left( -q^{0}\right) \delta \left(
q^{2}-m^{2}\right) \theta \left( q^{0}-p^{0}\right) \delta \lbrack (p-q)^{2}]%
\text{ .}  \label{jose}
\end{equation}
By virtue of the deltas and thetas, we have that $q^{2}=m^{2}$, $q^{0}<0$, $%
(p-q)^{2}=0$ and $q^{0}-p^{0}>0$. So, $p=(p-q)+q$ is the sum of two
4-vectors, one timelike and the other lightlike, both in the backward light
cone. So, $\ p$ is timelike. Then we can choose a Lorentz frame such that $\
p=(p^{0},\overrightarrow{0})$. We first consider the integral proportional
to $\ m$ in the integrand of this expression: 
\[
I_{1\mu }=\int dq\text{ }q^{\mu }\theta \left( -q^{0}\right) \theta \left(
q^{0}-p^{0}\right) \delta \left( q^{2}-m^{2}\right) \delta \lbrack (p-q)^{2}]%
\text{ .} 
\]
For $\mu \neq 0$ this integral vanishes by symmetry considerations. For $\mu
=0$ the integral can be solved in a straightforward manner and results 
\[
I_{10}=\frac{\pi }{4}p^{0}\left( 1-\frac{m^{4}}{\left( p^{0}\right) ^{4}}%
\right) \theta \lbrack (p^{0})^{2}-m^{2}]\theta \left( -p^{0}\right) \text{ .%
} 
\]
The corresponding term in (\ref{jose}), in any reference frame, is given by 
\begin{equation}
\widehat{d}_{-}(p)_{1}=\frac{1}{(4\pi )^{3}}p\!\!\!/\left( 1-\frac{m^{4}}{%
p^{4}}\right) \theta (p^{2}-m^{2})\theta \left( -p^{0}\right) \text{.}
\label{maria}
\end{equation}
Now we calculate the term proportional to $\slash\!\!\!q^{2}$ in (\ref{jose}%
): 
\[
I_{2\mu \nu }=\int dq\text{ }q_{\mu }q_{\nu }\theta \left( -q^{0}\right)
\theta \left( q^{0}-p^{0}\right) \delta \left( q^{2}-m^{2}\right) \delta
\lbrack (p-q)^{2}]\text{ .} 
\]
Also this integral vanishes for $\mu \neq \nu $ by symmetry reasons. By
Lorentz invariance it must be proportional to $g_{\mu \nu }.$ The
proportionality coefficient is determined by saturation with $g^{\mu \nu }$.
Thus 
\[
I_{2\mu \nu }=\frac{1}{4}g_{\mu \nu }\int dq\text{ }q^{2}\theta \left(
-q^{0}\right) \theta \left( q^{0}-p^{0}\right) \delta \left(
q^{2}-m^{2}\right) \delta \lbrack (p-q)^{2}]\text{ .} 
\]
This integral can now be easily solved and gets the following contribution
to integral (\ref{jose}): 
\begin{equation}
\widehat{d}_{-}(p)_{2}=\frac{m}{2(4\pi )^{3}}g_{\mu \nu }\beta ^{\mu }\beta
^{\nu }\left( 1-\frac{m^{2}}{p^{2}}\right) \theta (p^{2}-m^{2})\theta \left(
-p^{0}\right) \text{ .}  \label{joao}
\end{equation}
Combining (\ref{maria}) and (\ref{joao}) into (\ref{jose}) we get the final
result for $\widehat{d}_{-}(p)=\widehat{d}_{-}(p)_{1}+\widehat{d}_{-}(p)_{2}$
: 
\begin{equation}
\widehat{d}_{-}(p)=\frac{1}{(4\pi )^{3}}\theta (p^{2}-m^{2})\theta \left(
-p^{0}\right) \left( 1-\frac{m^{2}}{p^{2}}\right) \left[ \frac{m}{2}\beta
^{\mu }\beta _{\mu }+p\!\!\!/\left( 1+\frac{m^{2}}{p^{2}}\right) \right] 
\text{ .}  \label{d-fin}
\end{equation}
Turning back to (\ref{6}) we have 
\begin{eqnarray}
-e^{2}\beta ^{\mu }\widehat{d}_{-}(p)\beta _{\mu } &=&-\frac{e^{2}}{(4\pi
)^{3}}\theta (p^{2}-m^{2})\theta \left( -p^{0}\right) \left( 1-\frac{m^{2}}{%
p^{2}}\right) \left[ 2m+p\!\!\!/\left( 1+\frac{m^{2}}{p^{2}}\right) \right] 
\text{ }  \nonumber \\
&&  \label{mario} \\
&=&\widehat{r}_{\text{I}}^{\prime }(p)\text{,}  \nonumber
\end{eqnarray}
where we used the following algebraic relations for $\beta $ matrices 
\[
\beta ^{\mu }\beta ^{\nu }\beta _{\mu }=\beta ^{\nu } 
\]
and\footnote{%
We can show that this identity is valid using the usual scalar
representation of section 2. As the r.h.s. is invariant under changes of
representation, this result is general, i. e., representation independent.} 
\[
\beta ^{\mu }\beta ^{\nu }\beta _{\nu }\beta _{\mu }=4\text{ .} 
\]
The distribution $\widehat{d}_{+}(p)$ is calculated in the same way and
results 
\begin{equation}
-e^{2}\beta ^{\mu }\widehat{d}_{+}(p)\beta _{\mu }=\frac{e^{2}}{(4\pi )^{3}}%
\theta (p^{2}-m^{2})\theta \left( p^{0}\right) \left( 1-\frac{m^{2}}{p^{2}}%
\right) \left[ 2m+p\!\!\!/\left( 1+\frac{m^{2}}{p^{2}}\right) \right] \text{
.}  \label{mariana}
\end{equation}
Now, (\ref{mario}) and (\ref{mariana}) together give 
\begin{equation}
\widehat{d}_{\text{I}}(p)=\frac{e^{2}}{(4\pi )^{3}}\theta
(p^{2}-m^{2})\left( 1-\frac{m^{2}}{p^{2}}\right) \left[ 2m+p\!\!\!/\left( 1+%
\frac{m^{2}}{p^{2}}\right) \right] \text{sgn}(p^{0})\text{ .}  \label{disf}
\end{equation}

We verify that the singular order of this distribution is $\ \omega =1$. We
then split it by using the central splitting formula (\ref{cs}) with $\omega
=1$. For $p$ timelike this yelds 
\begin{equation}
\hat{r}_{\text{I}}(p^{0})=\frac{i}{2\pi }\left( p^{0}\right)
^{2}\int_{-\infty }^{+\infty }dk^{0}\frac{\hat{d}_{1}(k^{0})}{%
(k^{0}-i0)^{2}(p^{0}-k^{0}+i0)}  \label{28}
\end{equation}
We turn (\ref{28}) into this integral and solve it by standard methods \cite
{Scharf}. For an arbitrary Lorentz frame we arrive at 
\begin{eqnarray}
\hat{r}_{\text{I}}(p) &=&\frac{ie^{2}}{4(2\pi )^{4}}  \nonumber \\
&\times &\left\{ \left[ \text{log}\left| 1-b^{2}\right| -i\pi \text{sgn}%
\left( p^{0}\right) \theta \left( p^{2}-m^{2}\right) \right] \left[ m\left(
1-\frac{1}{b^{2}}\right) +\frac{p\!\!\!/}{2}\left( 1-\frac{1}{b^{4}}\right) %
\right] \right.  \nonumber \\
&-&\left. \frac{p\!\!\!/}{2b^{2}}-m-\frac{p\!\!\!/}{4}\right\} \text{ ,}
\label{30}
\end{eqnarray}
with $b^{2}=\frac{p^{2}}{m^{2}}$ . We can analitically extend this result
for arbitrary complex $p$: 
\begin{eqnarray}
\hat{r}_{\text{I}}^{\text{an}}(p) &=&\frac{ie^{2}}{4(2\pi )^{4}}\left\{ 
\text{log}\left( 1-b^{2}\right) \left[ m\left( 1-\frac{1}{b^{2}}\right) +%
\frac{p\!\!\!/}{2}\left( 1-\frac{1}{b^{4}}\right) \right] \right.  \nonumber
\\
&&\left. -\frac{p\!\!\!/}{2b^{2}}-m-\frac{p\!\!\!/}{4}\right\} \text{ ,}
\label{r1an}
\end{eqnarray}

Combining (\ref{30}) with (\ref{mario}), we obtain $T_{\text{I}}^{\text{self}%
}\left( x_{1},x_{2}\right) $: 
\begin{equation}
T_{\text{I}}^{\text{self}}\left( x_{1},x_{2}\right) =ie^{2}:\overline{\psi }%
(x_{1})\Sigma \left( x_{1}-x_{2}\right) \psi \left( x_{2}\right) :\text{ ,}
\label{Tsigma}
\end{equation}
where 
\begin{eqnarray}
\widehat{\Sigma }\left( p\right) &=&-i\left[ \widehat{r}_{\text{I}}(p)-%
\widehat{r}_{\text{I}}^{\prime }(p)\right]  \nonumber \\
&=&\frac{ie^{2}}{4(2\pi )^{4}}\left\{ \left[ \text{log}\left| 1-b^{2}\right|
-i\pi \theta \left( p^{2}-m^{2}\right) \right] \left[ m\left( 1-\frac{1}{%
b^{2}}\right) +\frac{p\!\!\!/}{2}\left( 1-\frac{1}{b^{4}}\right) \right]
\right.  \nonumber \\
&&\left. -\frac{p\!\!\!/}{2b^{2}}-m-\frac{p\!\!\!/}{4}\right\} \text{.}
\label{Sigma}
\end{eqnarray}
This result is free of infrared divergences, because we have used the
central splitting formula, which normalizes the solution at the point $p=0.$
Also, it is well defined on mass shell. Because $\omega =1$ we must add to
this result a first order polynimial in $p\!\!\!/$ in order to get the
general solution to the splitting problem. Thus, 
\begin{eqnarray}
\widetilde{\Sigma }\left( p\right) &=&\frac{ie^{2}}{4(2\pi )^{4}}\left\{ %
\left[ \text{log}\left| 1-b^{2}\right| -i\pi \theta \left(
p^{2}-m^{2}\right) \right] \left[ m\left( 1-\frac{1}{b^{2}}\right) +\frac{%
p\!\!\!/}{2}\left( 1-\frac{1}{b^{4}}\right) \right] \right.  \nonumber \\
&&\left. -\frac{p\!\!\!/}{2b^{2}}+C_{0}+C_{1}p\!\!\!/\right\} \text{.}
\label{Sigma gen}
\end{eqnarray}

We now consider radiative corrections due to the self energy insertions into
the DKP propagator. The complete propagator is 
\[
\widehat{T}_{\text{compl}}^{\text{c}}(p)=\widehat{T}^{\text{c}}(p)+\widehat{T%
}^{\text{c}}(p)\widetilde{\Sigma }\left( p\right) \widehat{T}^{\text{c}}(p)+%
\widehat{T}^{\text{c}}(p)\widetilde{\Sigma }\left( p\right) \widehat{T}^{%
\text{c}}(p)\widetilde{\Sigma }\left( p\right) \widehat{T}^{\text{c}}(p)+...%
\text{ ,} 
\]
where $\widehat{T}^{\text{c}}(p)$ is given by (\ref{prop}). Formally summing
this series we obtain 
\begin{eqnarray*}
\widehat{T}_{\text{compl}}^{\text{c}}(p) &=&\frac{\widehat{T}^{\text{c}}(p)}{%
1-\widetilde{\Sigma }\left( p\right) \widehat{T}^{\text{c}}(p)} \\
&=&\frac{1}{(2\pi )^{2}m}\frac{p\!\!\!/(p\!\!\!/+m)-p^{2}+m^{2}}{%
p^{2}-m^{2}+i0-\left[ p\!\!\!/(p\!\!\!/+m)-p^{2}+m^{2}\right] \widetilde{%
\Sigma }\left( p\right) }\text{ .}
\end{eqnarray*}
We require that $m$ be the physical mass of the DKP particle. This will be
satisfied only if 
\[
\left. \left[ p\!\!\!/(p\!\!\!/+m)-p^{2}+m^{2}\right] \widetilde{\Sigma }%
\left( p\right) \right| _{p^{2}=m^{2}}=0\text{.} 
\]
Substituting (\ref{Sigma gen}) into this condition we obtain the following
restriction on the arbitrary constants $C_{0}$ and $C_{1}$: 
\[
C_{0}=m\left( \frac{1}{2}-C_{1}\right) \text{ .} 
\]
This condition eliminates one of the arbitrary constants, say $C_{0}$. The
remaining one will be related by an Ward identity to another arbitrary
constant that will appear in the vertex distribution. In the Appendix we
derive this Ward identity by requiring gauge invariance of the theory.

Now we calculate the vertex distribution in the limit of zero tansferred
momentum by making use of the Ward identity (\ref{Wardder}) deduced in the
Appendix: 
\[
\widehat{\Lambda }^{\mu }\left( p,p\right) =\frac{1}{\left( 2\pi \right) ^{2}%
}\frac{\partial }{\partial p_{\mu }}\widehat{\Sigma }(p)\text{ .} 
\]
Substituting the explicit form of the self-energy (\ref{Sigma gen}) into
this identity we obtain 
\begin{eqnarray}
\widehat{\Lambda }^{\mu }\left( p,p\right) &=&\left. \frac{ie^{2}}{4(2\pi
)^{6}}\right\{ \left[ \text{log}\left| 1-b^{2}\right| -i\pi \theta \left(
p^{2}-m^{2}\right) \right]  \nonumber \\
&&\times \left[ \frac{2p^{\mu }}{mb^{4}}\left( 1+\frac{p\!\!\!/}{mb^{2}}%
\right) +\frac{\beta ^{\mu }}{2}\left( 1-\frac{1}{b^{4}}\right) \right] 
\nonumber \\
&&+\frac{2p^{\mu }}{m^{2}}\left[ \text{P.V.}\frac{1}{b^{2}-1}-i\pi \delta
(b^{2}-1)\right] \left[ m\left( 1-\frac{1}{b^{2}}\right) +\frac{p\!\!\!/}{2}%
\left( 1-\frac{1}{b^{4}}\right) \right]  \nonumber \\
&&\left. +\frac{p^{\mu }p\!\!\!/}{m^{2}b^{4}}-\frac{\beta ^{\mu }}{2b^{2}}%
+C_{1}\beta ^{\mu }\right\} \text{ .}  \label{vertex}
\end{eqnarray}
This result is well defined at $p=0$, but it is singular on mass shell $\
p^{2}=m^{2}$ due to the logarithmic term.

The above form of the vertex function suffices to study the physical meaning
of the constant $C_{1}$, which will be done in conection to charge
normalization. The physical charge is defined in the scattering of a scalar
particle by an external electromagnetic field at low energies. Thus we must
consider the contributions (in the limit of zero transferred momentum) to $S$
matrix from the terms containing $C_{1}$ in both self-energy and vertex
distributions. Because of the above mentioned mass shell singularity, we
must be care in taking the adiabatic limit. Making so, we can prove that all
these contributions cancel themselves and conclude that this constant has no
physical meaning. Nevertheless, it can be specified by requiring the vertex
function to satisfy the central splitting condition, i. e. 
\[
\widehat{\Lambda }^{\mu }\left( 0,0\right) =0\text{ .} 
\]
Using this condition into (\ref{vertex}) we obtain $C_{1}=\frac{1}{4}.$

\section{Concluding remarks}

In this paper we have considered scalar QED based on Duffin-Kemmer-Petiau
equation in the framework of Epstein-Glaser causal method. We have given the
basis to construct the second order S matrix and calculate the lowest order
distributions for Compton scattering, vacuum polarization and self-energy.
By using the gauge invariance requirement we determined the vertex
correction from the self-energy one.

The starting point of the causal approach was the identification of the
one-point distribution $T_{1}(x),$ which was given by the interaction term
in the Lagrangian (\ref{Lint}) of the theory, where all the fields entering
that expression were free fields.\ Thus, the causal method specified
completely the form of the interaction, giving us a non effective theory.
After determining, by using gauge invariance, the finite normalization
constant appearing in the two-point distribution for Compton scattering,
this distribution was identified with the propagator of the DKP scalar
field. Then, in the causal approach we have recovered, in a natural and even
simpler way, the basic quantities from which it is constructed the usual
effective theory. Namely, we identified $iT^{\text{c}}(x)$ as the effective
propagator and $-e:$ $\overline{\psi }(x)\beta ^{\mu }\psi (x):$ ($\psi $
and $\overline{\psi }$ being free fields) as the \ ``vertex'' (interaction).

At one loop level we have calculated the scalar vacuum polarization tensor,
the self-energy and the vertex correction. We have determined all physically
meaningful finite normalization constants from physical requirements as
symmetries and mass and charge normalization. Our results agree with that
obtained in the context of the effective theory. To do a complete analysis
at one loop level, we would have to calculate as well the singular order of
the four point distribution for scattering of two scalar particles. This
problem is fundamentally important to study the renormalizability of the
theory and is under our investigation presently. As future perspectives we
can quote the use of the causal approach to study DKP field interacting with
external gravitational fields.

\section{Acknowlegdments}

J.T.L. and B.M.P. thank respectively to CAPES-PICDT and CNPq \ for partial
support. L.A.M and J.S.V. thank FAPESP for full support.

\section*{Appendix: gauge invariance}

The gauge invariance requirement allows us to determine the remaining
arbitrary constant $C$, in the two-point distribution for Compton
scattering, and to find a relation (Ward identity) between the self-energy
and vertex distributions, thus relating the corresponding arbitrary
constants.

As we saw, in the causal approach all the fields are free fields. Then,
under a gauge transformation the electromagnetic field transforms as $A_{\mu
}(x)\rightarrow A_{\mu }(x)+\partial _{\mu }\Lambda (x)$, whereas the matter
fields $\psi (x)$ and $\overline{\psi }(x)$ remains unaffected. Here $%
\Lambda (x)$ is a c-number scalar field that satisfies $\square \Lambda
(x)=0 $ (Lorentz gauge) and vanishes at infinity.

The gauge invariance requirement amounts that in the adiabatic limit, $%
g\rightarrow 1$, the S matrix must be invariant under such transformations,
which implies the invariance of all the terms in the S matrix expansion (\ref
{sma}). We write a generic $n$-point distribution normally ordered with
respect to photon operators in the form 
\[
T_{n}(x_{1},...,x_{n})=\sum_{l=0}^{n}\sum_{1\leq k_{1}<...<k_{l}\leq
n}t_{k_{1}...k_{l}}^{\mu _{1}...\mu _{l}}(x_{1},...,x_{n}):A_{\mu
_{1}}(x_{k_{1}})...A_{\mu _{l}}(x_{k_{l}}):\text{ ,} 
\]
where $t_{k_{1}...k_{l}}^{\mu _{1}...\mu _{l}}(x_{1},...,x_{n})$ contains
only scalar operators and it is the sum of all graphs of order $n$ with only 
$l$ external photon lines, attached at the vertices $x_{k_{1}},...,x_{k_{l}}$%
. The external scalar operators are arbitrary. Applying the gauge invariance
requirement we arrive at the following condition \cite{Scharf}: 
\begin{equation}
\frac{\partial }{\partial x_{k_{j}}^{\mu _{j}}}t_{k_{1}...k_{l}}^{\mu
_{1}...\mu _{l}}(x_{1},...,x_{n})=0\text{ ,}  \label{gaugecond}
\end{equation}
for all $1\leq l\leq n$, all $1\leq j\leq l$, all $1\leq k_{1}<...<k_{l}\leq
n$ and all $(x_{1},...,x_{n})\in {\rm {\bf R^{4n}.}}$

When we apply this condition for the two-point Compton distribution, we get
at the condition 
\begin{equation}
\partial _{\mu }^{1}Q^{\mu \nu }(x_{1},x_{2})=0,  \label{gi}
\end{equation}
where 
\begin{eqnarray}
Q^{\mu \nu }(x_{1},x_{2}) &=&:\overline{\psi }(x_{1})\beta ^{\mu }t^{\text{I}%
}(x_{1},x_{2})\beta ^{\nu }\psi (x_{2}):  \nonumber \\
&&+:\overline{\psi }(x_{2})\beta ^{\nu }t^{\text{II}}(x_{1},x_{2})\beta
^{\mu }\psi (x_{1}):  \nonumber \\
&=&Q^{\nu \mu }(x_{2},x_{1})\text{ .}  \label{qmn}
\end{eqnarray}
and $\partial _{\mu }^{1}$ means derivative with respect to $x_{1}.$

\bigskip Turning (\ref{qmn}), (\ref{A}) and (\ref{B}) into this last
equation, the l.h.s. gives 
\begin{eqnarray}
\partial _{\mu }^{1}Q^{\mu \nu }(x_{1},x_{2}) &=&-:\partial _{\mu }\overline{%
\psi }(x_{1})\beta ^{\mu }[S^{\text{F}}(x_{1}-x_{2})-C\delta
(x_{1}-x_{2})]\beta ^{\nu }\psi (x_{2}):  \nonumber \\
&&-:\overline{\psi }(x_{1})\beta ^{\mu }[\partial _{\mu }S^{\text{F}%
}(x_{1}-x_{2})-C\partial _{\mu }\delta (x_{1}-x_{2})]\beta ^{\nu }\psi
(x_{2}):  \nonumber \\
&&-:\overline{\psi }(x_{2})\beta ^{\nu }[-\partial _{\mu }S^{\text{F}%
}(x_{2}-x_{1})+C\partial _{\mu }\delta (x_{2}-x_{1})]\beta ^{\mu }\psi
(x_{1}):  \nonumber \\
&&-:\overline{\psi }(x_{2})\beta ^{\nu }[S^{\text{F}}(x_{2}-x_{1})-C\delta
(x_{2}-x_{1})]\beta ^{\mu }\partial _{\mu }\psi (x_{1}):\text{ .}
\label{aaa}
\end{eqnarray}
From the equation of scalar Feynman propagator 
\[
(\square +m^{2})\triangle ^{\text{F}}(x)=\delta (x)\text{ ,} 
\]
the relation (\ref{vinculo}) and definition (\ref{A'}), we have 
\begin{equation}
i\partial \!\!\!/S^{\text{F}}(x)=mS^{\text{F}}(x)+\frac{i}{m}\partial
\!\!\!/\delta (x)\text{ .}  \label{propS}
\end{equation}
Using this into (\ref{aaa}), condition (\ref{gi}) yields 
\begin{equation}
C=\frac{I}{m}\text{ ,}  \label{AC}
\end{equation}
where $I$ $\ $is the $5\times 5$ identity matrix.

Now we consider the three-point distributions corresponding to the vertex
correction. We must consider four graphs, that are analogous to that we find
in the vertex correction of spinor QED. Considering the photon line attached
at the vertex $x_{3}$, the distribution $\ t_{3}^{\mu }(x_{1},x_{2},x_{3})$
we must enter into condition (\ref{gaugecond}) is 
\begin{eqnarray*}
t_{3}^{\mu }(x_{1},x_{2},x_{3}) &=&-:\overline{\psi }(x_{1})\Lambda ^{\mu
}(x_{1}-x_{3},x_{2}-x_{3})\psi (x_{2}): \\
&&+:\overline{\psi }(x_{1})\Sigma (x_{1}-x_{2})T^{\text{c}%
}(x_{2}-x_{3})\beta ^{\mu }\psi (x_{3}): \\
&&+:\overline{\psi }(x_{3})\beta ^{\mu }T^{\text{c}}(x_{3}-x_{1})\Sigma
(x_{1}-x_{2})\psi (x_{2}): \\
&&+:\overline{\psi }(x_{1})\beta ^{\rho }\psi (x_{1}):g_{\rho \nu }D_{0}^{%
\text{F}}(x_{1}-x_{2})\Pi ^{\nu \mu }(x_{2}-x_{3}) \\
&&+x_{1}\longleftrightarrow x_{2}\text{ ,}
\end{eqnarray*}
where $T^{\text{c}}$ is the \ ``effective DKP propagator'' given in (\ref{ep}%
). Thus we have 
\begin{eqnarray}
\partial _{\mu }^{3}t_{3}^{\mu }(x_{1},x_{2},x_{3}) &=&:\overline{\psi }%
(x_{1})\left[ -\partial _{\mu }^{3}\Lambda ^{\mu }(x_{1}-x_{3},x_{2}-x_{3})%
\right] \psi (x_{2}):  \label{ref} \\
&&+:\overline{\psi }(x_{1})\Sigma (x_{1}-x_{2})\partial _{\mu }^{3}\left[ T^{%
\text{c}}(x_{2}-x_{3})\beta ^{\mu }\psi (x_{3})\right] :  \nonumber \\
&&+:\partial _{\mu }^{3}\left[ \overline{\psi }(x_{3})\beta ^{\mu }T^{\text{c%
}}(x_{3}-x_{1})\right] \Sigma (x_{1}-x_{2})\psi (x_{2}):  \nonumber \\
&&+:\overline{\psi }(x_{1})\beta ^{\rho }\psi (x_{1}):g_{\rho \nu }D_{0}^{%
\text{F}}(x_{1}-x_{2})\left[ \partial _{\mu }^{3}\Pi ^{\nu \mu }(x_{2}-x_{3})%
\right]   \nonumber \\
&&+x_{1}\longleftrightarrow x_{2}\text{ .}  \nonumber
\end{eqnarray}
For the brackets in the second and third lines we have 
\begin{eqnarray*}
\partial _{\mu }^{3}\left[ T^{\text{c}}(x_{2}-x_{3})\beta ^{\mu }\psi (x_{3})%
\right]  &=&i\delta (x_{2}-x_{3})\psi (x_{3})\text{ ;} \\
\partial _{\mu }^{3}\left[ \overline{\psi }(x_{3})\beta ^{\mu }T^{\text{c}%
}(x_{3}-x_{1})\right]  &=&-i\overline{\psi }(x_{3})\delta (x_{3}-x_{1})\text{
,}
\end{eqnarray*}
where we have used the DKP equation and the fact that $T^{\text{c}}$ is the
corresponding DKP Green function (see eq. (\ref{GF})). Turning these results
into (\ref{ref}) and taking into account the delta distributions, we obtain 
\begin{eqnarray}
\partial _{\mu }^{3}t_{3}^{\mu }(x_{1},x_{2},x_{3}) &=&:\overline{\psi }%
(x_{1})\left\{ -\partial _{\mu }^{3}\Lambda ^{\mu
}(x_{1}-x_{3},x_{2}-x_{3})\right.   \nonumber \\
&&\left. +i\delta (x_{2}-x_{3})\Sigma (x_{1}-x_{2})-i\delta
(x_{1}-x_{3})\Sigma (x_{1}-x_{2})\right\} \psi (x_{2}):  \nonumber \\
&=&0\text{ .}
\end{eqnarray}

This condition will be satisfied only if the term into curl brackets vanish.
Thus we arrive to the desired Ward identity: 
\begin{equation}
-\partial _{\mu }^{3}\Lambda ^{\mu }(x_{1}-x_{3},x_{2}-x_{3})+i\delta
(x_{2}-x_{3})\Sigma (x_{1}-x_{2})-i\delta (x_{1}-x_{3})\Sigma (x_{1}-x_{2})=0%
\text{ .}  \label{Ward}
\end{equation}
Denoting 
\[
y_{1}=x_{1}-x_{3}\text{\quad and\quad }y_{2}=x_{2}-x_{3}\text{,} 
\]
we can write 
\[
\frac{\partial }{\partial x_{3}^{\mu }}\Lambda ^{\mu }\left(
y_{1},y_{2}\right) =-(\partial _{1}+\partial _{2})_{\mu }\Lambda ^{\mu
}\left( y_{1},y_{2}\right) 
\]
and equation (\ref{Ward}) becomes 
\[
(\partial _{1}+\partial _{2})_{\mu }\Lambda ^{\mu }\left( y_{1},y_{2}\right)
+i\delta (y_{2})\Sigma (y_{1}-y_{2})-i\delta (y_{1})\Sigma (y_{1}-y_{2})=0%
\text{ .} 
\]
Taking the Fourier transform\footnote{%
Our convention to the Fourier transform of a distribution $F(y_{1},y_{2})$
is 
\[
\widehat{F}(p_{1},p_{2})=\frac{1}{(2\pi )^{4}}\int dy_{1}dy_{2}\text{ e}%
^{i(p_{1}y_{1}+p_{2}y_{2})}F(y_{1},y_{2}) 
\]
} of this equation we arrive at 
\[
(p_{1}+p_{2})_{\mu }\widehat{\Lambda }^{\mu }\left( p_{1},p_{2}\right) -%
\frac{1}{\left( 2\pi \right) ^{2}}\left\{ \widehat{\Sigma }(p_{1})-\widehat{%
\Sigma }(p_{2})\right\} =0\text{ .} 
\]
Introducing the notation $q=-p_{2},$ $p=p_{1}$ and redefining $\widehat{%
\Lambda }^{\mu }$ as 
\[
\widehat{\Lambda }^{\mu }(p,-q)\rightarrow \widehat{\Lambda }^{\mu }(p,q), 
\]
\bigskip\ we arrive at 
\begin{equation}
(p-q)_{\mu }\widehat{\Lambda }^{\mu }\left( p,q\right) =\frac{1}{\left( 2\pi
\right) ^{2}}\left\{ \widehat{\Sigma }(p)-\widehat{\Sigma }(q)\right\} \text{
.}  \label{17}
\end{equation}
Taking the limit $p\rightarrow q$ we have 
\begin{equation}
\widehat{\Lambda }^{\mu }\left( p,p\right) =\frac{1}{\left( 2\pi \right) ^{2}%
}\frac{\partial }{\partial p_{\mu }}\widehat{\Sigma }(p)\text{ ,}
\label{Wardder}
\end{equation}
which is the form of the Ward identity we use to determine the vertex
correction in Section {\bf 7}.


\begin{thebibliography}{99}
\bibitem{Itzikson}  C. Itzykson and J. B. Zuber, \ {\it Quantum Field Theory
(}Mc-GrawHill, Singapore, 1980).

\bibitem{Petiau}  G. Petiau, {\it Acad. R. Belg. Cl. Sci. M\'{e}m. Collect 8}
{\bf 16, }No. 2 (1936).

\bibitem{Kemmer 1}  N. Kemmer, {\it Proc. Roy. Soc. }{\bf A 166 }(1938) 127.

\bibitem{Duffin}  R. J. Duffin, \ {\it Phys. Rev. }{\bf 54 }(1938) 1114.

\bibitem{Kemmer 2}  N. Kemmer, {\it Proc. Roy. Soc. }{\bf A 173} (1939) 91.

\bibitem{Umezawa}  H. Umezawa, {\it Quantum Field Theory} (North-Holland,
Amsterdan, 1956).

\bibitem{Bogoliubov}  N. N. Bogoliubov and D. V. Shirkov, {\it Introduction
to the Theory of Quantized Fields (}Wiley, New York, 1980).

\bibitem{Pimentel 2}  J. T. Lunardi, B. M.Pimentel, R. G. Teixeira and J. S.
Valverde, {\it Phys. Lett. }{\bf A 268 }(2000) 165.

\bibitem{Berestetskii}  A. I. Akhiezer and V. B. Berestetskii, {\it Quantum
Electrodynamics} (Interscience, New York, 1965).

\bibitem{Krajcik}  R. A. Krajcik and M. M. Nieto, {\it Am. J. Phys.} {\bf 45}
(1977) 818.

\bibitem{Pimentel 3}  V. Ya. Fainberg and B. M. Pimentel. {\it Phys. Lett. }%
{\bf A 271 }(2000) 16.

\bibitem{prl}  B. C. Clark, S. Hama, G. R. K\"{a}lbermann, R. L. Mercer and
L. Ray, {\it Phys. Rev. Lett. }{\bf 55 }(1985) 592.

\bibitem{prc1}  E. Friedman, G. K\"{a}lberman and C. J. Batty, {\it Phys.
Rev.} {\bf C} {\bf 34 }(1986) 2244.

\bibitem{Nowakowski}  M. Nowakowski, {\it Phys. Lett.} {\bf A 244} (1998)
329.

\bibitem{DKP curvo}  J. T. Lunardi, B. M. Pimentel and R. G. Teixeira,
``Duffin-Kemmer-Petiau equation in Riemannian Space-times'' in ``{\it %
Proceedings of Workshop on Geometrical Aspects of Quantum Fields},'' A. A.
Bytsenko, A. E. Gon\c{c}alves and B. M. Pimentel (eds.), (World Scientific,
Singapore, 2001), p. 111.

\bibitem{Pimentel 1}  V. Ya. Fainberg and B. M. Pimentel, {\it Theor. Math.
Phys}. {\bf 124} (2000) 445.

\bibitem{Gribov}  V. Gribov, {\it Eur. Phys. J.} {\bf C 10 }(1999) 71.

\bibitem{prog}  L. K. Kerr, B. C. Clark, S. Hama, L. Ray, and G. W Hoffmann, 
{\it Prog. Theor. Phys.} {\bf 103} (2000) 321.

\bibitem{ScharfNC}  M. Dutsch, F. Krahe and G. Scharf, {\it Il Nuovo Cim. }%
{\bf A 106 }(1993) 277.

\bibitem{Epstein}  H. Epstein and V. Glaser, {\it Ann. Inst. H. Poincar\'{e} 
}{\bf A} {\bf 19} (1973) 211.

\bibitem{Scharf}  G. Scharf, {\it Finite Quantum Electrodynamics: the Causal
Approach}, 2$^{{\rm nd}}$ ed. (Springer, Berlin, 1995).

\bibitem{irreps}  J. G\'{e}h\'{e}niau, {\it Acad. R. Belg. Cl. Sci. M\'{e}m.
Collect 8 }{\bf 18, }No. 1 (1938).

\bibitem{Aste}  A. Aste, {\it Ann. Phys. }{\bf 257} (1997) 158.

\bibitem{ScharPim}  G. Scharf, W. F. Wreszinski, B. M. Pimentel and J. L.
Tomazelli, {\it Ann. Phys.} {\bf 231} (1994) 185.

\bibitem{mpt}  L. A. Manzoni, B. M. Pimentel and J. L. Tomazelli, {\it Eur.
Phys. J. }{\bf C8} (1999) 353; {\it ibid.} {\bf 12} (2000) 701.

\bibitem{trace}  T. Kinoshita, {\it Progr. Theor. Phys. }{\bf 3} (1950) 473.

Harish-Chandra, {\it Proc. Roy. Soc.} {\bf A 186} (1946) 502.
\end{thebibliography}
\end{document}